\documentclass[aps,prl,twocolumn,groupedaddress]{revtex4}
\usepackage{amssymb}
\usepackage{graphicx}
\begin{document}

\title{Punctuated Equilibrium in Software Evolution}
\author{A.A.Gorshenev, Yu.M.Pis'mak}
\affiliation{Department of Theoretical Physics State University of
Saint-Petersburg, Ul'yanovskaya 1, Petrodvorets, 198904
Saint-Petersburg, Russia}
\date{\today}

\begin{abstract}
The approach based on paradigm of self-organized criticality
proposed for experimental investigation and theoretical modelling
of software evolution. The dynamics of modifications studied for
three free, open source programs Mozilla, Free-BSD and Emacs using
the data from version control systems. Scaling laws typical for
the self-organization criticality found. The model of software
evolution presenting the natural selection principle is proposed.
The results of numerical and analytical investigation of the model
are presented. They are in a good agreement with the data collected for
the real-world software.
\end{abstract}

\maketitle

The basic self-organization mechanisms of complex systems
in the Nature are intensively studied last years.
The proposed in the pioneering paper of P.Bak, C.Thang
and K.Wiesenfeld \cite{BTW}  paradigm of self-organized
criticality (SOC) appeared to be most fruitful here.
The SOC dynamics is characterized by
avalanche-like changes of the system state with power law
statistics of the avalanche growth.
The main feature of the SOC regime is that it is an attractor of the
system dynamics approached without any fine tuning of control
parameters.

Studies of the fossil records have shown that the
biological evolution is a strong non-equilibrium process with
long periods of stasis interrupted by avalanches of large changes in
biosphere . This is a main point of the
punctuated equilibrium conception of biological evolution
suggested by E.Gould and H.Eldridg \cite{GE, GE1}. Detailed
quantitative analysis of paleontological dates  revealed  the scaling
power laws in distributions of avalanches in extinction
and creations of species \cite{RS, RB}. Therefore the biological evolution can
be considered as a kind of SOC dynamics. This has been demonstrated by P.Bak
and K.Sneppen in the proposed model of Darwinian selection in ecosystem
\cite{baksneppen}. The development of computer science and engineering
created the "virtual biosphere" with  specific evolution laws
of "virtual species" -- computer programs.
In this paper we propose an approach to the studies of software
evolution in the framework of the SOC conception.

"Life" of large computer program is a perfect example of
evolutionary process in complex system. During its creation the
program often undergoes multiple internal reorganizations.
New devices and platforms supported, new features added,
system tuning performed, erroneous code corrected, huge
number of cosmetic changes going on during the development of any
program \cite{critique, decay}.
Despite of the fact
that the first papers on software evolution study are now decades
old, the universal mechanisms of computer program evolution are
unclear. The most of existing in this region research methodics are
based on assumption that estimations of possible changes
in a program can be obtained without taking into account
underlying dynamical laws creating this system \cite{opensource,
release}. In the multitude of papers the authors propose  statistical
methods predicting the number of defects in a program
using of some kind of metrics describing complexity, size, volume
etc. \cite{akiyama, halstead}.

From our point of view the main disadvantage of such approach is
that even the best in the world static metric which forecasts a
number of improvements to be done in computer program to
correspond a given specification, becomes useless if the
specification changes in time essentially. Our approach can be
considered as an elaboration of a prototype for dynamical metrics
based on the use of characteristics of SOC universality class of
the system.

There is a lot of phenomenological work has been done on
software evolution. Lehman's laws suggest
that as system grows in size, it becomes increasingly difficult to
add new code unless explicit steps are taken to reorganize the
overall design \cite{nineties, lehmanold}. There were some systems
examined both at system level and within the top-level subsystems.
It has been noted that subsystems can behave quite differently
from the system as whole\cite{release, opensource} Good
metaphors such as ``code decay'' has been proposed to describe the
continuous process that makes the software more brittle over
time\cite{decay, parnas}. Thus the software evolution  has many
similar features with the evolution of biological species, and one can
expect that evolution of large computer program  presents some
class of universality of the SOC dynamics.

To study software evolution processes it is necessary to have
information about the state of the system in different moments of
time. The usual sources
 of such data  are various versions or
releases of a product \cite{opensource, nineties, release}.
Unfortunately, the number of releases rarely exceeds  a couple of
tens. This fact significantly decreases our possibility to study
the evolution of program. The better sources of information
about changes in computer programs are version control systems.
One of them is Concurrent Versions System (CVS). It keeps
information about changes happened in short time intervals.

Using the CVS in our work, we studied the histories of three
software projects: Mozilla web browser, Free-BSD Operating System
and Gnu Emacs text editor \cite{freebsd, mozilla, emacs}. For each
of these projects we analyzed only files written in the basic for
the project language. These are: C++ for Mozilla , C for Free-BSD
and Lisp for Emacs. Header files for C/C++ were not studied. Total
amounts of the processed files  are approximately 9000,
11000, 900  for Mozilla, Free-BSD, Emacs . Total lengths of
RCS-files are $1\cdot10^7$,$1\cdot10^7$ and $2\cdot10^6$ lines.
Total amount of the data processed  exceeds 2 Gigabytes. Due to
some resource limitations only part of the Free-BSD CVS storage
processed. Histories of all three projects are stored under
control of the CVS and were publicly available during our research
period from the corresponding Internet servers \cite{freebsd,
mozilla, emacs}.

\begin{figure}
\includegraphics[totalheight=8cm, angle=-90]{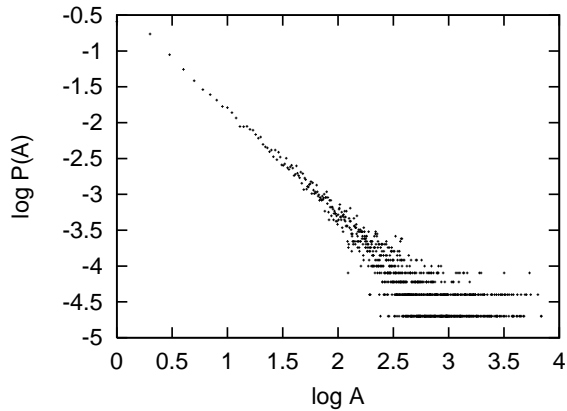}
\caption{Distribution P(A) for Free-BSD.}
\label{fig:freebsd_addition}
\end{figure}

\begin{figure}
\includegraphics[totalheight=8cm, angle=-90]{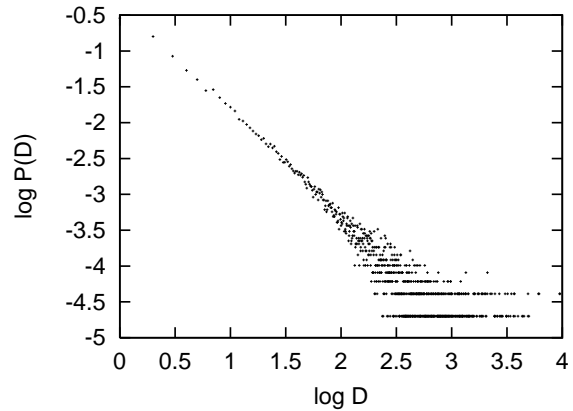}
\caption{Distribution P(D) for Free-BSD.}
\label{fig:freebsd_deletion}
\end{figure}

For each change of each file an amount $D$ of deleted lines and an
amount $A$ of  added lines  were collected. Empty lines and
comments were collected together with the rest of the data. A
number of lines in the very first version of each file was not
counted. Distributions $P(A)$ and $P(D)$ were evaluated for these
two arrays $A_i$ and $D_i$. As an example the data for the
Free-BSD are shown in the FIG.~\ref{fig:freebsd_addition} and
FIG.~\ref{fig:freebsd_deletion}
in log-log scale.
Results for the Emacs and Mozilla are similar. One can see
that power functions are accurate
approximations for $P(A)$ and $P(D)$:  $P(A) \sim A^{\mu_a}, P(D)
\sim D^{\mu_d}$.
The values of exponents are the following:\\
\begin{tabular}{rrr}
Free-BSD:&$\mu_a=-1.44\pm 0.02,$&$\mu_d=-1.48\pm0.02$\\
Mozilla:&$\mu_a=-1.43\pm 0.02,$&$\mu_d=-1.47\pm0.02$\\
Emacs:&$\mu_a=-1.39\pm 0.03,$&$\mu_d=-1.49\pm0.04$\\
\end{tabular}
These scaling laws can be considered as a manifestation of the SOC
in the evolution  of software.

One of the important notion being used in description of the SOC
dynamic is the avalanche.  The SOC process  can be presented as a
consequence of meta-stable states interrupted by the
avalanche-like changes in the system. For evolution of computer
program  the close analog of the avalanche is a set of changes
going on from version to version. We see that the avalanche
statistic in evolution of software is described by power functions
with nontrivial exponents. The universality of SOC dynamical mechanisms
allows one to hope that a simple "holistic" model  can be constructed for
its quantitative description \cite{Bak}. To realize this idea
for software evolution modeling we use the following assumptions.
 The specific of software changes is that one
programmer can not modify a program at different points
simultaneously (at least using a traditional development tools).
The point of changes is characterized as "weakest" one in the
program text: a programmer has some subjective estimation of parts
of a program and makes changes in
place which is estimated as  extremely non-satisfactory. If the
change is made on some point, corresponding changes must be made
in some other places, i.e. in the program there is a coordination
structure of its elements.
We suppose that changes in the program
can't make its size less that some minimal one.

  We formulate the model presenting this conception
as follows. Computer program as a system constitutes a sequence of
elements -- lines of code. At time point $t$ the $i-th$ line is
characterized by a number $b_i(t)$, $0<b_i(t)<1$ representing its
"fitness" in the program text or a barrier in respect to change in
future stages of evolution. The state of the system of $N$
elements is fully given  by the set of barriers $B(t)=\{b_i(t),
i=1, 2, ..., N\}$. The evolution of the program is described in our
model as a sequence of  $B(t)$ for discrete time points  $t=0,1,2,...$.
The coordination structure of program is
presented by a network of its elements, where each element-node is
conformed with its nears neighbors. The node having minimal
barrier is defined as weakest unit of the system.
At each time point $t$ we define the set $W(t)$ containing weakest unit with
all its neighbors. We call $W(t)$ the weakest spoil at time $t$.

Dynamics in the model is defined in the following way. The initial
number of nodes $N (0)$ and the minimal possible number of
nodes $K$ are supposed to be given. The initial values of barriers
$b_i(0)$ are chosen at random. The state $B(t)$ at time point $t$
transforms into state $B(t+1)$ as follows. If the number
of nodes $N(t)$ in the system is more than $K$ two kinds of
changes are possible. With probability $\alpha$, the weakest unit
is deleted from the system or with probability $1-\alpha$ a new neighbor
node to the weakest unit is inserted into the system.
After that the barriers of all nodes from weakest spoil $W(t)$ are sat random.
So if $N(t)>K$, the size of the system
decreases or increases by one for one time step. If $N(t)=K$, then
deletion is impossible and the above described insertion is made.

  Our model is a modification of well known Simple Model
of Biological Evolution suggested by Bak and Sneppen
\cite{baksneppen}, and its essential specific is that the number of
system elements variates in time. In our study we have considered
two versions of the model: with 1-dimensional (1D) and random
neighbor (RN) coordination structure. In the 1D case the nodes are
organized into 1D lattice with periodic boundary condition, and
each node has two neighbors. In RN model, there is no fixed
coordination structure in the system, and at each time step
$k$ random nodes are chosen as neighbors of weakest unit. We have
considered the case $k=1$ only.

An avalanche as the elementary process of complex behavior of
non-equilibrium dynamical system can be defined in different ways.
Usually in the model of SOC dynamics the $\lambda$- and transient
avalanches are considered \cite{baksneppen, maslov, hierarchy}. In
studies of our model we were interested mostly in transient
avalanches. They can be defined as follows. Let at the time  moment
$t_0$ the minimal barrier has the value $f_0$.
The sequence of $S$ time steps during which the minimal barrier
does not exceed $f_0$: $b_{min}(t)< f_0, t_0 < t < t_0+S$ is
called transient avalanche or just avalanche if it finishes at the
time point $t_0+S$ when the value of minimal barrier becomes larger than
$f_0$: $b_{min}(t+S)> f_0$. Distribution $P(S)$ of avalanche
temporal duration and distribution $P(R)$ of avalanche spatial
volume are important characteristics of the type of dynamics.  For
our model it is reasonable to consider two values as
characteristics of volume of changes produced in the system  by
avalanche. One of them is the number $A$  of new elements appeared
in the system at the end of avalanche. Other is the number $D$ of
elements disappeared from the system at the end of the avalanche.
In dynamic of our model we studied mostly the distributions $P(S),
P(A), P(D)$ of temporal and spatial characteristics of avalanches.

 We studied numerically the 1D and RN versions of the model for
$\alpha=\frac{1}{2}$. The initial size of the system was $8000$
elements. The experiment went on until one million of avalanches
were registered. We got the following results.
The $P(S)$, $P(A)$, $P(D)$ distributions can be
sufficiently approximated by the power functions $P(S) \sim
S^{\tau}$, $P(A) \sim A^{\mu_a}$, $P(D) \sim D^{\mu_d}$  with
exponents $\tau=-1.358\pm0.005$, $\mu_a-1.45\pm0.01$,
$\mu_d=-1.47\pm0.02$ for the 1D model and $\tau=-1.901\pm0.008$,
$\mu_a-1.98\pm0.01$, $\mu_d=-2.10\pm0.02$ for the RN model.

 For the RN model it is possible to obtain analytical description
in the framework of master equation formalism. To do it one can
use the method of construction of master equation proposed for
analysis of the SOC dynamic of random neighbor version of
Bak-Sneppen model \cite{FSB}. If we denote $P_{n,N}(t)$ the
probability that at time point $t$ there are N nodes in the
system, and $n$ of ones have barriers less than $\lambda$, where
$0<\lambda<1$, then the dynamical rules of RN model result in the
following master equation
\begin{equation}
P_{n,N}(t+1) = (\alpha+\beta\delta_{N,K+1})P^{a}_{n, N}(t)+ \beta
P^{d}_{n, N}(t)
\end{equation}
where $\beta = 1-\alpha$, and in terms of $\mu=1-\lambda$,
$\rho_{n,N}=(n-1)/(N-1)$, $\sigma_{n,N}=1-\rho_{n,N}$ the
quantities $P^{d}_{n,N}(t)$ and $P^{a}_{n,N}(t)$ can be presented
by the following relations:
\begin{widetext}
$$
\begin{array}{rrl}
P^{a}_{n,N}(t)&=&A_{n+2,N-1}^aP_{n+2,N-1}(t)+B_{n+1,N-1}^aP_{n+1,N-1}(t)+
C_{n,N-1}^aP_{n,N-1}(t)
+D_{n-1,N-1}^a  P_{n-1,N-1}(t)+\\
    &&+E_{n-2,N-1}^aP_{n-2,N-1}(t)+
    (\mu^3\delta_{n, 0} + 3\lambda\mu^2\delta_{n,1} + 3\lambda^2\mu\delta_{n, 2} +
    \lambda^3\delta_{n, 3})P_{0, N-1}(t),\\
P^{d}_{n,N}(t)&=&A_{n+2,N+1}^dP_{n+2,N+1}(t)+B_{n+1,N+1}^dP_{n+1,N+1}(t)+
    C_{n,N+1}^dP_{n,N+1}(t)+ (\mu\delta_{n,
    0}+\lambda\delta_{n,1})P_{0,N+1}(t),\\
A^a_{n, N}&=&\mu^3\rho_{n,N},\ 
B^a_{n, N}=3\lambda\mu^2\rho_{n,N}+\mu^3\sigma_{n,N},\ 
C^a_{n, N}=3\lambda\mu^2\sigma_{n,N}+3\lambda^2\mu\rho_{n,N},\ 
D^a_{n, N}=3\lambda^2\mu\sigma_{n,N}+\lambda^3\rho_{n,N},\\
E^a_{n, N}&=&\lambda^3\sigma_{n,N},\ 
A^d_{n, N}=\mu\rho_{n,N},\ 
B^d_{n, N}=\mu\sigma_{n,N}+\lambda\rho_{n,N},\ 
C^d_{n, N}=\lambda\sigma_{n,N}.
\end{array}
$$
\end{widetext}
Here, the coefficients $A^a_{n, N}$, $B^a_{n, N}$, $C^a_{n,N}$,
$D^a_{n, N}$, $E^a_{n, N}$, $A^d_{n, N}$, $B^d_{n, N}$,
$E^d_{n,N}$ are defined in the last two lines for $0<n\leq N$. For
$n\leq 0$ and $n>N $ they assumed to be zero. The master equation
(1) enables one to find $P_{n,N}(t)$ for $t>0$, if initial values
$P_{n,N}(0)$ are given. Basing on this equation one can obtain
analytical results for characteristics of dynamics in RN model.
With that end in view it is convenient to use the formalism of
generating function appeared to be very effective for construction
of exact solution for master equations of RN version of
Bak-Sneppen model
  \cite{pismak,pismak1,pismak2, pismak3}. Dynamic of RN model
is more complex than one of Bak-Sneppen model, and solution of
master equation for RN model for software evolution appears to be
not easy problem. Here, we present only the exact result for
$P_N(t)= \sum_{n=0}^N P_{n,N}(t)$ being the probability that the
system has $n$ element with barriers less then $\lambda$ at time
point $t$. Let us denote ${\cal N}(y,u)$ the generating function
for probabilities $P_N(t)$: ${\cal N}(y,u)= \sum_{N=K,t=0}^\infty
P_N(t)y^{N-K}u^t$. From (1) we obtain the following equation for
${\cal N}(y,u)$:
$$
  {\cal N}(y,u)[y-u(\alpha y^2+\beta)] = y {\cal
N}(y,0)+
    u\beta[y^2-1]{\cal N}(0,u).
$$
It describes  1-dimensional discrete diffusion with reflection
and can be solved by methods, used in \cite{pismak,pismak1,pismak2, pismak3}
. The result has the form
$$
 {\cal N}(y,u)=\frac{y {\cal N}(y,0)
(u-\tau)+ u\alpha\tau {\cal N}(\tau,0)(y^2-1) }{
(u-\tau)[y-u(\alpha y^2+\beta)]}
$$
where $\tau=(1-\sqrt{1-4u^2\alpha\beta})/2u\alpha$ is the
analytical in the point
 $u=0$  solution of equation
$[\tau-u(\alpha \tau^2+\beta)]=0$. Mean value $n(t)= K +
\sum_{N}P_N(t)N$ of the system element number at time point $t$
has the following asymptotic for large $t$: $n(t)\approx
(2\alpha-1)t$ for $\alpha>1/2$, $n(t)\approx \sqrt{2t/\pi}$ for
$\alpha= 1/2$, and if we denote $p_{ev}$ ($p_{od}$ ) the
probability that the initial number $N(0)$ of nodes is even (odd),
then
$$
n(t)\approx
K+[1+(-1)^{t+K}(1-2\alpha)^{2}(p_{ev}-p_{od})]/[2(1-2\alpha)]
$$
for $\alpha < 1/2$. The corrections to the leading terms of
asymptotic are of the form : $ n(0) -
\frac{2\alpha\beta}{(\alpha^2-\beta^2)} + f(t)g_1(t)$, for $\alpha
> 1/2 $, $t^{-1/2}g_3(t)$ for $\alpha = 1/2$ and $ f(t)g_2(t)$,
for $\alpha < 1/2 $. Here $f(t)=[4\alpha\beta]^{t/2}/t^{-3/2}$ and
$g_i(t) $, $i=1,2,3$ are bounded for large t , i.e. there are
constants $T$, $M$ that $|g_i(t)|<M$, if $t>T$. Since
$4\alpha\beta<1$ for $\alpha\neq1/2$, the function $f(t)$
decreases exponentially fast for large $t$.

The asymptotic  behavior of $n(t)$ demonstrates the dynamical
phase transition at the point $\alpha = 1/2$ . For
$\alpha<1/2$ the volume of system remains finite , but for
$\alpha\geq 1/2$ it can became as large as one likes. At the point
$\alpha=1/2$, the dynamics of the system is critical one.

In above formulated model we tried to present elementary
mechanisms of software changes.  They are made by programmer
locally in the place where these changes most of all needed. But
a program changed in one place often must be
changed in other places in some way connected to the first one.
For example, in order to change the number of arguments of
subroutine call, one needs to change not only the line containing
the call operator but the definition of the subroutine either.
This would lead to some subsequent changes of all the calls to the
subroutine in all the program. If one adds the line in which some
data read from a disk one should add some lines to check whether
the data have been read successfully, and this in turn can require
some change in the list of the modules included which in turn can
cause a name conflict which in turn can cause other changes, etc..
Thus, the avalanche-like processes seems to be natural for
modifications of programs. Avalanche ends up when all the
parts of the program code are more or less satisfy some subjective
and implicit criteria of programmer.
Naively speaking,  the program as whole becomes "a little
bit better".  In the model it can be presented  as a
process terminating when the value of minimal barrier becomes greater than
initial one.
This was the point why we studied
transient avalanches of self organization period and not the
$\lambda$-avalanches of the stationary mode. The obtained
statistical characteristics of avalanches make it possible to
conclude that SOC is the dominating dynamical regime in evolution
of free software. Our results demonstrate that the natural
selection can create this type of "punctuated equilibrium"  
of such complex "virtual beings" in info-sphere. We
believe that in the framework of proposed approach the modern
methods of investigation of the
SOC dynamics can appear to be very effective for studies of
basic problem of software evolution. Our results could be seen
also as a theoretical prerequisite for the development of new tools and
methods for advanced measures of software quality engineering.

The work of Yu.M. Pis'mak was supported in part by  Russian
Foundation   for   Basic   Research  (Grant  No 03-01-00837),
Swiss National Science Foundation (SCOPES Grant  No 7SUPJ062295)
and Nordic Grant for Network Cooperation with
the Baltic Countries and Northwest Russia No FIN-6/2002

\bibliographystyle{unsrt}
\bibliography{Literat.bib}

\end{document}